# Interactive Evolution of Multiple Water-Ice Reservoirs on Mars: Insights from Hydrogen Isotope Compositions

Hiroyuki Kurokawa[a,b*], Tomohiro Usui[c], Masahiko Sato[d,e]

[a]*Earth-Life Science Institute, Tokyo Institute of Technology, 2-12-1 Ookayama, Meguro-ku, Tokyo 152-8550, Japan*

[b]*Department of Physics, Nagoya University, Furo-cho, Chikusa-ku, Nagoya, Aichi 464-8602, Japan*

[c]*Department of Earth and Planetary Sciences, Tokyo Institute of Technology, 2-12-1 Ookayama, Meguro-ku, Tokyo 152-8551, Japan*

[d]*Geological Survey of Japan, National Institute of Advanced Industrial Science and Technology, Central 7, 1-1-1 Higashi, Tsukuba, Ibaraki 305-8567, Japan*

[e]*Department of Environmental Changes, Kyushu University, 744 Motooka, Nishi-ku, Fukuoka 819-0395, Japan*

## Abstract

Remote sensing data from orbiter missions have proposed that ground ice may currently exist on Mars, although the volume is still uncertain. Recent analyses of Martian meteorites have suggested that the water reservoirs have at least three distinct hydrogen isotope compositions (D/H ratios): primordial and high D/H ratios, which are approximately the same and six times that of ocean water on Earth, respectively, and a newly identified intermediate D/H ratio, which is approximately two to three times higher than that in ocean water on Earth. We calculate the evolution of the D/H ratios and the volumes of the water reservoirs on Mars by modeling the exchange of hydrogen isotopes between multiple water reservoirs and the atmospheric escape. The D/H ratio is slightly higher in the topmost thin surface-ice layer than that in the atmosphere because of isotopic fractionation by sublimation, whereas the water-ice reservoir just below the exchangeable topmost surface layer retains the intermediate D/H signature found in Martian meteorites.

* Corresponding author (email: hiro.kurokawa@elsi.jp)

We propose two possible models for constraining the volume of the ground ice considering the observed D/H ratios and geomorphological estimates of Paleo-oceans. The first assumes that the atmospheric loss is dominated by the Jeans escape. In this case, the volume of ground ice should be larger than the total volume of the observable surface ice that mainly occurs as polar layered ice deposits. The other model assumes diffusion-limited atmospheric loss in which the interactive evolution of the multiple water reservoirs naturally accounts for the observed D/H ratios. In this scenario, a large volume of ground ice does not necessarily exist currently on Mars as opposed to the perspective view proposed on the basis of recent orbiter missions.

*Keywords:* Mars, water, ice, isotope, atmospheric escape

# 1. Introduction

Mars currently has a cold and dry surface environment, and relatively small amounts of water ice have been observed at the polar caps (Jakosky & Phillips, 2001; Christensen, 2006). The cold conditions on Mars have likely occurred since the end of the Noachian (~3.5 Ga, Hartmann & Neukum, 2001), as indicated by the decreased rates of erosion caused by running surface water and the formation of U-shaped valleys caused by ice since that time (Jakosky & Phillips, 2001). However, geomorphological evidence such as ancient valley networks and deltas requires repeated episodes of liquid water runoff in the Noachian period (e.g., Hynek et al., 2010; Di Achille & Hynek, 2010). Moreover, the Hesperian outflow channels suggest a large total volume of water (e.g., Carr & Head, 2015).

The current reservoir of surface water occurs mainly as polar layered deposits (PLD), although the total volume of this reservoir is significantly smaller than the estimated volume of the Paleo-oceans (e.g., Kurokawa et al., 2014). An increasing amount of evidence obtained by the Mars Odyssey Gamma Ray Spectrometer (Boynton et al., 2002, 2007) and the Mars Express radar sounder (Mouginot et al., 2012) suggests the presence of additional reservoirs of water in the form of ground ice at middle to high latitudes. The presence of recurring slope lineae (McEwen et al., 2014) in a wide range of regions including low-latitudes implies that water exchange is ongoing between the atmosphere and the underground cryosphere.

The hydrogen isotope ratio (D/H) in a water reservoir reflects the evolutionary history of the water in the reservoir. Kurokawa et al. (2014) suggested that the volume of the subsurface ice might exceed that of the PLD. The evolution of the total volume of the water reservoirs was determined on the basis of the D/H ratios found in Martian meteorites, whereby the Martian water reservoirs were treated as a single reservoir with a uniform D/H ratio. However, multiple near-surface water reservoirs with different D/H ratios have been recently been proposed (Fig. 1 and Table 1). As a result of isotope fractionation during the atmospheric escape, a relatively high mean δD ($\delta D = [(D/H)_{sample}/(D/H)_{reference} - 1] \times$ 1000, using standard mean ocean water as a reference) value (δD ≃ 5000‰) has been determined for water in the Martian atmosphere through telescope observations (Owen et al., 1988; Krasnopolsky et al., 1997; Novak et al., 2011; Villanueva et al., 2015; Krasnopolsky, 2015; Aoki et al., 2015) and through Mars rover explorations (Webster et al., 2013). Analyses of hydrogen isotopes in geochemically enriched Martian meteorites (Shergotty, Los Angeles, LAR 06319, and GRV 020090) have shown that surficial water reservoirs have similarly high δD values, indicating that exchange occurs between the water in the reservoirs and the atmosphere (Greenwood et al., 2008; Usui et al., 2012; Hu et al., 2014). Intermediate δD values (δD ≃ 1000–2000‰) have been detected in surficial water reservoirs through ion microprobe analyses of matrix glasses in Martian meteorites (Usui et al., 2015). These intermediate δD values are distinct from the high δD values detected in magmatic water. A similar intermediate δD value (δD ≃ 2000‰) was detected in 3.5 Ga mudstone recently analyzed by the *Curiosity* rover (Mahaffy et al., 2015). Usui et al. (2015) suggested that the reservoir of water with an intermediate D/H ratio could be ground ice or hydrated crust distributed throughout Mars. These measurements indicate that different surface and subsurface water reservoirs on Mars currently have different D/H ratios caused by the various evolutionary histories of the reservoirs. Strong local anisotropies and seasonal variability of the atmospheric D/H ratio (Villanueva et al., 2015) may support the presence of multiple water reservoirs with different D/H ratios.

Exchange among multiple water reservoirs has been assessed in some theoretical studies. Montmessin et al. (2005) used three-dimensional global circulation calculations, considering isotope fractionation in sublimation processes, to demonstrate that the D/H ratio in the atmosphere is approximately 15% lower than that in the permanent ice cap. Fisher (2007) showed that the D/H ratio is higher in atmospheric water vapor than that in the surface ice because of fractionation during atmospheric escape and sublimation. The results of these studies and the increasing amount of evidence for the presence of multiple water reservoirs with different D/H ratios suggest the importance of studying the evolution of Martian water, taking into consideration the existence of multiple water reservoirs. In the present study, we examine the evolution of the D/H ratios and volumes of Martian water reservoirs during the post-Noachian cold stages (3.5 Ga to the present) by using a new multi-water-reservoir model that uses geochemical constraints determined from the hydrogen isotope compositions found for the identified Martian water reservoirs.

We explain the numerical multi-water-reservoir model and provide analytical solutions for steady state conditions in Section 2. In Section 3, by using the steady state solution, we assess the D/H ratios of current Martian water-ice reservoirs. The evolution of the D/H ratios is calculated, and the current volumes of water-ice reservoirs on Mars are estimated in Section 4. Finally, the evolution of the D/H ratios and the volumes of water reservoirs through Martian history is discussed in Section 5.

# 2. Model

## *2.1. Formulation*

We calculated the long-term evolution of the Martian water reservoirs by using a multi-reservoir-box model based on the model of Fisher (2007). The evolution of the D/H ratios and the volumes of water in the atmosphere and water ice were calculated in our model (Fig. 2). Seasonal and local variations were neglected for simplification. Because we focused on long-term evolution based on atmospheric escape, we treated annually and globally averaged states in this study. The ice was divided into the topmost exchangeable ice and unexchangeable ice. Water is photolyzed into hydrogen and oxygen and is lost from the atmosphere through atmospheric escape at an escape rate $\Phi$. Water is exchanged between the exchangeable ice and the atmosphere by sublimation (i.e., from the ice to the atmosphere) and desublimation (i.e., from the atmosphere to the ice); both processes together are called simply as sublimation. The sublimation rates are $\varepsilon_{\mathrm{v}}$ for atmosphere to ice transfer and $\varepsilon_{\mathrm{i,e}}$ for ice to atmosphere transfer. Water is supplied to the exchangeable ice from the unexchangeable ice at the supply rate $\varepsilon_{\mathrm{i}}$. This supply rate does not necessarily denote the actual supply rate but includes the permeation of the topmost ice into the subsurface ice. Molecular diffusion, hydrothermal activity, volcanism, and impacts are possible mechanisms of the permeation. The isotopic fractionation of hydrogen is considered by using fractionation factors for atmospheric escape $f$ and sublimation from the atmosphere to the ice $\alpha$.

The time derivatives of the volumes of water vapor $q_{\mathrm{v}}$, exchangeable water ice $q_{\mathrm{i,e}}$, and unexchangeable water ice $q_{\mathrm{i}}$ are given by

$$\frac{\mathrm{d}q_{\mathrm{v}}}{\mathrm{d}t} = -\Phi + \varepsilon_{\mathrm{i,e}} - \varepsilon_{\mathrm{v}}, (1)$$

$$\frac{\mathrm{d}q_{\mathrm{i,e}}}{\mathrm{d}t} = -\varepsilon_{\mathrm{i,e}} + \varepsilon_{\mathrm{v}} + \varepsilon_{\mathrm{i}}, (2)$$

and

$$\frac{\mathrm{d}q_{\mathrm{i}}}{\mathrm{d}t} = -\varepsilon_{\mathrm{i}}. (3)$$

The time derivatives of the D/H ratios of the water vapor $a_{\mathrm{v}}$, exchangeable water ice $a_{\mathrm{i,e}}$, and unexchangeable water ice $a_{\mathrm{i}}$ are given by

$$\frac{\mathrm{d}}{\mathrm{d}t}(q_{\mathrm{v}}a_{\mathrm{v}}) = -f\Phi a_{\mathrm{v}} + \varepsilon_{\mathrm{i,e}}a_{\mathrm{i,e}} - \alpha\varepsilon_{\mathrm{v}}a_{\mathrm{v}}, (4)$$

$$\frac{\mathrm{d}}{\mathrm{d}t}(q_{\mathrm{i,e}}a_{\mathrm{i,e}}) = -\varepsilon_{\mathrm{i,e}}a_{\mathrm{i,e}} + \alpha\varepsilon_{\mathrm{v}}a_{\mathrm{v}} + \varepsilon_{\mathrm{i}}a_{\mathrm{i}}, (5)$$

and

$$\frac{\mathrm{d}a_{\mathrm{i}}}{\mathrm{d}t} = 0. (6)$$

The δD values of water vapor, exchangeable water ice, and unexchangeable water ice are represented by $\delta D_V$, $\delta D_{i,e}$, and $\delta D_i$, respectively. It should be noted that our treatment of deuterium exchange shown in Eqs. 4 and 5 is based on Eq. 16 in a study by Fisher (2007) with some modifications. The volume of water vapor $q_V$ was determined from the saturated vapor pressure of water. Thus, volume $q_V$ is expected to be constant. We can therefore assume that the right side of Eq. 1 is zero, i.e.,

$$\Phi = \varepsilon_{i,e} - \varepsilon_V. (7)$$

The volume of exchangeable ice $q_{i,e}$ is determined from the permeable thickness of $H_2O$ molecules in the atmosphere, which depends on a permeation mechanism such as molecular diffusion, hydrothermal activity, volcanism, and impacts. The thickness was assumed to be constant in our model. Thus, in Eq. 2,

$$\varepsilon_{i,e} = \varepsilon_V + \varepsilon_i. (8)$$

The water reservoir evolution was determined by using Eqs. 1–8 in Section 4.

## *2.2. Steady state solutions*

As will be shown in Sections 3 and 4, the D/H ratio of the water vapor $a_V$ is in a steady state with the D/H ratio of the exchangeable ice, which changes over time. In the steady state, the right side of Eq. 4 is zero, i.e.,

$$-f\Phi a_V + \varepsilon_{i,e}a_{i,e} - \alpha\varepsilon_V a_V = 0. (9)$$

Combining Eqs. 7 and 9 allows the relationship between the D/H ratios of the atmosphere and the exchangeable ice to be given by

$$\frac{a_{\mathrm{v}}}{a_{\mathrm{i,e}}} = \frac{(\varepsilon_{\mathrm{i,e}}/\Phi)}{\alpha(\varepsilon_{\mathrm{i,e}}/\Phi) - (\alpha - f)}. (10)$$

The timescale $\tau_V$ required to realize Eq. 10 is

$$\tau_{\mathrm{v}} \sim \frac{a_{\mathrm{v}}}{\mathrm{d}a_{\mathrm{v}}/\mathrm{d}t} = \frac{q_{\mathrm{v}}}{(\alpha - f)\Phi + (\frac{a_{\mathrm{i,e}}}{a_{\mathrm{v}}} - \alpha)\varepsilon_{\mathrm{i,e}}}. (11)$$

The timescale required is substantially shorter than the geological timescale because of the small volume of water vapor $q_V$ in which the water column height is ~10 μm. The D/H ratio of the exchangeable ice $a_{i,e}$ will be assessed by using Eq. 10 in Section 3.

The D/H ratio of the exchangeable ice $a_{i,e}$ also has an asymptotic steady state. Assuming that the time derivatives of both $a_V$ and $a_{i,e}$ in Eqs. 4 and 5 are zero, the steady state is given by

$$\frac{a_{i,e}}{a_i} = \frac{\alpha(\varepsilon_{i,e}/\Phi) - (\alpha - f)}{f(\varepsilon_{i,e}/\Phi)}. \quad (12)$$

Although the steady state relationship between the atmospheric D/H ratio and the exchangeable ice D/H ratio (Eq. 10) is obtained in a short time, that between the D/H ratios of exchangeable and unexchangeable ice (Eq. 12) requires a considerably longer time for establishment because the volume of the exchangeable ice is orders of magnitude greater than the volume of water vapor. Therefore, a steady state is not always reached between the exchangeable and unexchangeable ice. The evolution of the D/H ratio of the exchangeable ice is analyzed using Eq. 12 in Section 4.

## *2.3. Volumes and fluxes*

We assumed that the precipitable water column height $q_V$ is 10 μm (Smith, 2002; Fisher, 2007). Although $q_V$ should vary with the planetary obliquity that chaotically oscillates over geological time (Fisher, 2007; Laskar et al., 2004), it does not affect the relationship between the D/H ratios of atmospheric and exchangeable water reservoirs under the steady state condition (Eq. 10).

The volume of the exchangeable ice $q_{i,e}$ depends on the permeation mechanism; thus, it was treated as a free parameter in our model. For example, considering the diffusion of semi-heavy water (HDO) molecules in water ice, where the diffusion coefficient is $\sim 10^{-14}$ m$^2$ s$^{-1}$ (Kuhn & Thurkauf, 1958; Montmessin et al., 2005), the estimated permeable thickness is ~10 m in 1 Gyr. The permeable thickness can be larger if another permeation mechanism such as hydrothermal activity, volcanism, or impact is included. We used $q_{i,e}$ = 10 m global equivalent layer (GEL) as a reference value, where 1 m GEL corresponds to $1.4 \times 10^{14}$ m$^3$ (Kurokawa et al., 2014). These assumptions give the maximum permeable thickness for molecular diffusion. The volume of the unexchangeable ice $q_i$ did not explicitly appear in our calculations, although it should be sufficiently large for balancing the loss of exchangeable ice.

Water vapor in the atmosphere can become photochemically dissociated to escape as hydrogen and oxygen; the amount of hydrogen escaping is assumed to be regulated by the loss of oxygen (Liu & Donahue, 1976). Lammer et al. (2003a) estimated that the total amount of oxygen lost through atmospheric escape over the last 3.5 Gyr is equivalent to 24–58 m GEL water (14 − 34 m GEL in the original paper, in which a different conversion factor is used; see Kurokawa et al., 2014, for details). We used the total loss of 24 m GEL as a reference value. The time dependence of the escape rate was taken from the exponential fit of the escape rate determined by Lammer et al. (2003a), which is given by $\Phi \propto \exp[-1.646 \times (t/1\ \mathrm{Gyr})]$, where $t$ is time. This value was considered to be the minimum estimate because the oxygen produced through the dissociation of water vapor can also be consumed through the oxidation of surface material (Lammer et al., 2003b; Kurokawa et al., 2014). The minimum escape rate is later referred to as $\Phi = \Phi_{\mathrm{min}}(t)$. The sublimation rate $\varepsilon_{\mathrm{i,e}}$ is estimated in Section 3, and the sublimation rate of $\varepsilon_{\mathrm{i,e}} = \Phi + 10\ \mu\mathrm{m\ yr}^{-1}$ is used in Section 4.

## *2.4. Fractionation factors*

The fractionation factor for atmospheric escape $f$ depends on time because the escape regime and rate change over time. For the current Martian condition, $f$ was estimated to be 0.016 according to a photochemical model calibrated by observations (Krasnopolsky et al., 1998; Krasnopolsky, 2000). However, fractionation factors in ancient times are poorly constrained. Thus, we estimated the value of $f$ as 3.5 Ga by following the procedures described below.

The net fractionation factor $f$ is the product of two fractionation processes: fractionation caused by H and D having different escape rates and that between $H_2$ and $H_2O$. The latter has been estimated to be 0.41 (Krasnopolsky & Feldman, 2001). The former fractionation factor depends on the escape regime of hydrogen. The hydrogen escape is dictated by a combination of the escape rate from the exobase to space and the diffusion rate from the homopause to the exobase. If the diffusion rate is higher than the escape rate, the escape rate limits the hydrogen loss. This regime typically occurs under current Martian conditions (e.g., Lammer et al., 2008). However, if the escape rate is higher than the diffusion rate, the diffusion rate limits the hydrogen loss. Hydrogen loss from Earth is currently limited by this mechanism (e.g., Greenberg et al., 1993). The escape regime of hydrogen in the ancient Martian condition is uncertain because it depends on the solar ultraviolet (UV) and Martian atmospheric conditions at the time such that a higher UV flux results in a higher escape rate, whereas a higher hydrogen mixing ratio at the homopause results in a higher diffusion rate. Therefore, we estimated the fractionation factor between H and D during the atmospheric escape for the two escape regimes of Jeans escape-limited and diffusion-limited.

First, we considered the case in which hydrogen loss is limited by the Jeans escape rate from the exobase. The Jeans escape formula (Jeans, 1925) is given by

$$w_{\mathrm{J}} = (1+\lambda)e^{-\lambda}\left(\frac{k_{\mathrm{B}}T_{\mathrm{exo}}}{2\pi m}\right)^{\frac{1}{2}}, (13)$$

where $\lambda = m\,g\,r_{\mathrm{exo}}/(k_{\mathrm{B}}\,T_{\mathrm{exo}})$, $w_{\mathrm{J}}$ is the Jeans velocity, $k_{\mathrm{B}}$ is the Boltzmann constant, $m$ is the mass of the particle, $g$ is the gravitational acceleration at the exobase, $r_{\mathrm{exo}}$ is the distance of the exobase from the center of the planet, and $T_{\mathrm{exo}}$ is the exospheric temperature. Given that $T_{\mathrm{exo}}$ is $\simeq$ 800 K and $r_{\mathrm{exo}}$ is $\simeq r_{\mathrm{Mars}}$ + 800 km (calculated for 3.8 Ga, Tian et al., 2009, where $r_{\mathrm{Mars}}$ is the Martian radius), the former fractionation factor is estimated to be $w_{\mathrm{J}}(\mathrm{D})/w_{\mathrm{J}}(\mathrm{H}) \simeq 0.26$. Therefore, the net fractionation factor $f$ is $\simeq 0.26 \times 0.41 \simeq 0.11$ at 3.8 Ga. Tian et al. (2009) did not provide $T_{\mathrm{exo}}$ and $r_{\mathrm{exo}}$ at 3.5 Ga; therefore, we used $f$ = 0.11 as an estimate for the fractionation factor at 3.5 Ga in our model. It should be noted that the net fractionation factor for current Martian conditions ($f$ = 0.016) was not derived from Eq. 13 because nonthermal escape rather than the Jeans escape of atomic deuterium is the dominant mechanism of the deuterium loss for such conditions (Krasnopolsky et al., 1998). Although the nonthermal escape rate of deuterium is poorly understood for early Mars, thermal escape would be more efficient than that under current Martian conditions. Therefore, we considered $f$ = 0.016–0.11 for the Jeans escape-limited case. A detailed modeling of the upper atmosphere and nonthermal escape is needed for more accurate estimation of the fractionation factor.

Second, we estimated the fractionation factor in the case in which the hydrogen loss was limited by the diffusion rate of hydrogen from the homopause to the exobase. The fractionation factor for the diffusion-limited escape $f_{\mathrm{dif}}$ of species 1 and 2 is given by

$$f_{\mathrm{dif}} = \frac{b_2(m_{\mathrm{a}} - m_2)}{b_1(m_{\mathrm{a}} - m_1)}, (14)$$

where $m_{\mathrm{a}}$ is the atmospheric mean mass, $m_{\mathrm{i}}$ is the mass of species $i$, and $b_{\mathrm{i}}$ is the binary diffusion parameter for $i$ in atmosphere (Mandt et al., 2009). The ratio of the binary diffusion coefficients was estimated as (e.g., Genda & Ikoma, 2008)

$$\frac{b_{\mathrm{D}}}{b_{\mathrm{H}}} = \sqrt{\left(\frac{m_{\mathrm{a}} + m_{\mathrm{D}}}{m_{\mathrm{a}} \times m_{\mathrm{D}}}\right) / \left(\frac{m_{\mathrm{a}} + m_{\mathrm{H}}}{m_{\mathrm{a}} \times m_{\mathrm{H}}}\right)}. (15)$$

The net fractionation can be $f \simeq 0.41 \times 0.70 \simeq 0.29$ for H and D or $0.41 \times 0.81 \simeq 0.33$ for $H_2$ and HD. We considered $f = 0.29$–$0.33$ and adapted $f = 0.33$ as the maximum estimate for fractionation induced by diffusion-limited escape. In summary, we used the fractionation factors for hydrogen loss of $f = 0.016$, with $f = 0.11$ for the Jeans escape-limited case, and $f = 0.33$ for the diffusion-limited case in the following sections.

The fractionation factor for sublimation $\alpha$ depends on the sublimation processes occurring. Direct deposition does not lead to fractionation ($\alpha = 1$), whereas $\alpha$ for equilibrium frosting and precipitation has been estimated to be 1.35 (Fisher et al., 2008). Three-dimensional global circulation calculations have shown that the atmospheric D/H ratio is 15% smaller than the ice D/H ratio (Montmessin et al., 2005), which corresponds to a mean fractionation factor $\alpha \simeq 1/0.85 \simeq 1.18$. Therefore, we used the range $\alpha = 1$–$1.35$ for the sublimation fractionation factor in our model.

## 3. D/H ratios of the atmospheric and exchangeable ice reservoirs under steady state conditions

In this section, we assess the D/H ratio of the exchangeable ice $a_{i,e}$, which is related to the observed D/H ratio of the Martian atmosphere $a_V$ through sublimation, by using our steady state model (Eq. 10). By using the model, $a_V/a_{i,e}$ was found to negatively correlate with $\varepsilon_{i,e}/\Phi$ (Fig. 3). The asymptotic value of $a_V/a_{i,e}$ is $\alpha^{-1}$ for a large $\varepsilon_{i,e}/\Phi$, as shown by Eq. 10. When $\varepsilon_{i,e}/\Phi = 1$, Eq. 10 gives $a_V/a_{i,e} = f^{-1}$, which agrees with the analytical solution for the atmospheric isotope ratio determined from the balance between supply and escape (Yung et al., 1988; Jakosky et al., 1994).

The actual value of $\varepsilon_{i,e}/\Phi$ is estimated in two different ways: from the observed changes in the column density of the water vapor and from the saturated vapor pressure of water. As Fisher (2007) stated, the annual change in the precipitable water column height ($\sim$10 μm yr$^{-1}$) gives an estimated sublimation rate of $\varepsilon_{i,e} \sim 10^{16}$ m$^{-2}$ s$^{-1}$. Given the escape rate of hydrogen at $\Phi = 1.6 \times 10^{12}$ m$^{-2}$ s$^{-1}$ (Yung et al., 1988), we estimated $\varepsilon_{i,e}/\Phi \sim 10^4$. However, the saturated vapor pressure of water $p_{SW}$ at the mean diurnal temperature at the poles, $T = 150$ K (Carr, 2007), was determined to be $p_{SW} \sim 10^{-5}$ Pa (Murphy & Koop, 2005). The sublimation rate was estimated to be $\varepsilon_{i,e} \sim p_{SW}/(m_{H2O}\, v_{th}) \sim p_{SW}/(m_{H2O}\, k_B\, T)^{1/2} \sim 10^{18}$ m$^{-2}$ s$^{-1}$, where $v_{th}$ is the thermal velocity of the water vapor. This gives $\varepsilon_{i,e}/\Phi \sim 10^6$. Therefore, our two different approaches provided a probable $\varepsilon_{i,e}/\Phi$ value range of $\sim 10^{4-6}$. These estimated values are sufficiently high for resulting in $a_v/a_{i,e}$ $\sim \alpha^{-1}$.

The $a_v/a_{i,e}$ ratio is governed by fractionation during sublimation rather than that during escape, which means that the D/H ratio in the exchangeable ice is higher than that in the water vapor in the atmosphere. The calculated value is $a_v/a_{i,e} \simeq \alpha^{-1} \simeq 0.7$–1 ($\alpha$ = 1–1.35). This steady state is reached in a short time, $\tau_v \sim q_v/\varepsilon_{i,e} < 1$ yr (Eq. 11). Thus, the Martian ice able to be exchanged with the atmosphere is expected to have a D/H ratio of α times that of the atmospheric D/H ratio.

By using the observed mean $\delta D_v \simeq 5000$‰ shown in Fig. 1 (Owen et al., 1988; Novak et al., 2011; Webster et al., 2013), our steady-state model gave $\delta D_{i,e} \simeq 5000$–7100‰, which is consistent with the maximum δD value of δD ≃ 6000‰ found in Martian meteorites (Usui et al., 2012; Hu et al., 2014) as shown in Fig. 1 and Table 1. This result supports the hypothesis that the high δD signatures in Martian meteorites reflect the existence of exchangeable water reservoirs on Mars (e.g., Watson et al., 1994; Boctor et al., 2003). Hydrogen isotope analyses performed by the *Curiosity* rover using the tunable laser spectrometer of the Sample Analysis at Mars investigation suite also showed that the D/H ratio is slightly higher in the Rocknest fines, at δD = 5880‰ than in the atmosphere, at δD = 4950‰, Webster et al., 2013) as shown in Fig. 1. This result is consistent with sublimation equilibrium occurrence.

The reservoir of the intermediate D/H ratio (δD = 1000–2000‰; Fig. 1 and Table 1) detected in glassy phases in Martian meteorites (Usui et al., 2015) is not water ice exchangeable with the atmosphere because $\delta D_{i,e}$ should be higher than $\delta D_v$ (≃5000‰) in such ice. Thus, this reservoir could be unexchangeable ice buried under the topmost exchangeable ice (Fig. 2).

# 4. Evolution of water reservoirs during the post-Noachian cold stages (3.5 Ga to the present)

## *4.1. Constraints from D/H analyses of meteorites, telescope observations, and rover explorations*

In this section, the evolution of the D/H ratios and volumes of the water reservoirs during the cold stages (3.5 Ga to the present) are calculated to assess the volume of the Martian ground ice. The evolution history is constrained from measurements of the D/H ratios of the water reservoirs (Fig. 1 and Table 1). The initial D/H ratio of the water in the near-surface reservoirs is identical to the D/H ratio of the Martian mantle (δD = 275‰; Usui et al., 2012), which was determined from melt inclusions found in the Martian meteorite Y-980459 (shergottite). Isotope analyses of the Martian meteorite ALH 84001, which has a crystallization age of 4.1 Gyr (Bouvier et al., 2009; Lapen et al., 2010), allowed the D/H ratio to be determined for 4.1 Ga (1200–3000‰, Boctor et al., 2003; Greenwood et al., 2008). This δD value could represent the δD value for the bulk of the Paleo-ocean because efficient mixing owing to a high degree of water mobility is expected to have occurred in that era (e.g., Hynek et al., 2010; Di Achille & Hynek, 2010). Once the Martian surface environment changed to its current cold state, the frozen near-surface water reservoirs likely had heterogeneous D/H ratios because of the reduced water mobility at low temperature. The exchangeable and unexchangeable ice likely evolved in different ways. The D/H ratio of the exchangeable ice is comparable to or slightly higher than the D/H ratio of the atmosphere (δD ≃ 5000‰, Owen et al., 1988; Novak et al., 2011; Webster et al., 2013) as mentioned in Section 3. Moreover, the unexchangeable ice could have an intermediate D/H ratio of 1000–2000‰ (Usui et al., 2015) as also mentioned in Section 3.

In our model, the initial δD values of the water vapor $\delta D_V$, the exchangeable ice $\delta D_{i,e}$, and the unexchangeable $\delta D_i$ were assumed to be 1000–2000‰. Although $\delta D_V$ and $\delta D_{i,e}$ evolved owing to atmospheric escape and sublimation, $\delta D_i$ remained constant over time. The present volumes of the water-ice reservoirs can satisfy these observed D/H ratios.

## *4.2. Evolution of the D/H ratios of near-surface water reservoirs*

D/H ratios of the water reservoirs are used to constrain the volume of the reservoirs. We first show the manner in which the D/H ratios evolve depending on the parameter ($f$, $\alpha$, $\Phi$, $q_{i,e}$). A typical case for the evolution of δD values with $f = 0.016$, $\alpha = 1.18$, and initial δD = 1000‰ is shown in Fig. 4. The δD values were found to increase more slowly over time because the escape rate $\Phi$ decreases over time. The steady state of $a_V/a_{i,e}$ was reached in a considerably shorter timescale than that required for the changes in $a_{i,e}$, as mentioned in Section 3. Thus, $a_V/a_{i,e}$ was determined to always be at the steady state (Eq. 10), with $a_{i,e}/a_V \simeq \alpha = 1.18$. As $\delta D_{i,e}$ increased in time, $\delta D_V$ also increased because of the steady state relationship. The $a_V/a_{i,e}$ ratio was at the steady state in all of our evolutionary calculations.

The dependence of $\delta D_{i,e}$ on the fractionation factors $f$ and $\alpha$ is shown in Fig. 5. The results depended moderately on $f$ and slightly on $\alpha$. A larger $f$ resulted in a smaller $\delta D_{i,e}$, as expected, owing to inefficient fractionation occurring through the atmospheric escape. For the parameter set shown in Fig. 5, the Jeans escape-limited cases ($f = 0.016$ and $f = 0.11$) gave $\delta D_{i,e}$ values comparable to or slightly higher than 5000‰. Although the actual volume of the exchangeable ice $q_{i,e}$ and the actual escape rate $\Phi$ were unknown, the D/H ratios calculated from the realistic parameter set were consistent with the δD range determined from analyses of Martian meteorites and from evolved water in Rocknest fines observed by the *Curiosity* rover (Usui et al., 2012; Webster et al., 2013) as shown in Fig. 1. Our model results were consistent with these constraints. On the contrary, the $\delta D_{i,e}$ value was lower than 5000‰ in the diffusion-limited case ($f = 0.33$).

The interactive evolution of multiple water reservoirs yielded an asymptotic upper limit of exchangeable ice D/H ratio. The evolution of $\delta D_{i,e}$ with an artificially increased escape rate $\Phi$ is shown in Fig. 6 for the diffusion-limited case ($f = 0.33$). The time evolution of $a_{i,e}$ showed asymptotic behavior as $\Phi$ increased. The asymptotic value corresponds to the steady state of $a_{i,e}$ with $a_i$ given by Eq. 12 and is approximated as $a_{i,e}/a_i \sim \alpha/f$. In Fig. 6, the upper limit is $\delta D_{i,e} \simeq 5000$‰ for $f = 0.33$ and $\alpha = 1$, which means that $\delta D_{i,e} = 5000$‰ might be realized without dependence on the escaped volume of water if the volume is sufficiently large. It should be noted that the $a_{i,e}$ value of the upper limit depends on parameters $a_i$, $f$, and $\alpha$. The upper limit is substantially higher in the Jeans escape-limited case ($f = 0.016$–$0.11$). If we assume $f = 0.11$ and $\alpha = 1.18$, the asymptotic value of $\delta D_{i,e}$ is $\simeq$20000‰.

## *4.3. Constraints on near-surface ice volume*

The volumes of the exchangeable and unexchangeable ice, $q_{i,e}$ and $q_i$, are constrained by using the D/H ratios. We determined that the results should satisfy $\delta D_{i,e} =$ 5000‰, considering the representative δD values detected for Martian meteorites (Fig. 1). The $\delta D_{i,e}$ value might be higher if the δD of $\simeq$6000‰, obtained from the evolved water in Rocknest fines by *Curiosity* (Webster et al., 2013), represents the typical isotopic composition of the water ice. Our model predicted that the $\delta D_{i,e}$ is comparable to or slightly higher than the $\delta D_v$ ($\simeq$5000‰). The actual value of $\delta D_{i,e}$ is uncertain; thus, we selected $\delta D_{i,e} = 5000$‰, which is a commonly used value.

The parameter space in which $\delta D_{i,e} = 5000$‰ was satisfied for $f = 0.016$, 0.11, and 0.33, was shown in Fig. 7. The range shown in each panel of the figure was caused mostly by uncertainty in the initial δD of 1000–2000‰. The D/H ratios of the atmosphere and exchangeable ice in the δD = 2000‰ case were 1.5 times higher than the initial δD value of 1000‰. This result means that the escaped volume of water required for $\delta D_{i,e} = 5000$‰ is 1/1.5 times smaller. However, $\alpha$ (= 1–1.35) was found to have little effect on the ranges (Fig. 5). A larger $q_{i,e}$, larger $f$, or smaller $\alpha$ requires a larger escaping volume $q_{loss}$. The required water loss for $f$ = 0.33 (Fig. 7c) does not have an upper limit because the asymptotic upper limit of $\delta D_{i,e}$ is 5000‰ for $\alpha = 1$ and the initial δD of 1000‰ (Fig. 6).

First, the Jeans escape-limited case ($f$ = 0.016–0.11) was considered (Fig. 7a and Fig. 7b). We determined $q_{i,e}$ and $q_i$ by comparing our results for the $f$ = 0.016 and $f$ = 0.11 cases with the geomorphological estimates for the Paleo-oceans. Estimates of the oceanic volume vary from 100 m GEL to 690 m GEL (Head et al., 1999; Carr & Head, 2003; Di Achille & Hynek, 2010). The topographies providing these estimates have various crater-retention ages. It is uncertain whether this represents the evolution of the volume or that of the form (liquid water or solid ice) of the near-surface water reservoir. Here, we regarded the variation in the estimates to be the uncertainty in the volume of the Paleo-ocean.

The minimum estimated $q_{i,e}$ and $q_i$ values are given by the minimum estimated for the Paleo-ocean. We used 100 m GEL (Head et al., 1999) as the total volume ($q_{i,e} + q_i + q_{loss}$) in Figs. 7a and 7b. The vertical axis is $q_{loss}$ as a function of $q_{i,e}$. The difference between the volume of the Paleo-ocean and $q_{loss}$ corresponds to $q_i$. Using $q_{i,e}$ = 10 m GEL, the permeable volume for molecular diffusion, gave $q_{loss} \simeq$ 20–26 m GEL for $f$ = 0.016–0.11. In this case, unexchangeable ice with $q_i \simeq$ 64–70 m GEL should be present. The total volume of ice, $q_{i,e} + q_i \simeq$ 74–80 m GEL, is then larger than the current volume of the detected PLD of 20–30 m GEL (Zuber et al., 1998; Plaut et al., 2007). The difference, 44–60 m GEL, may correspond to the volume of the ground ice. If we assume that $q_{i,e}$ is $\simeq$30 m GEL, $q_{loss}$ is $\simeq$ 70 m GEL, in which case there would currently be no unexchangeable ice. The exchangeable volume of $\simeq$ 30 m GEL is comparable to the total volume of the PLD. Not enough molecular diffusion occurs for the volume predicted, and thus another permeation mechanism is required.

Substantially larger estimates of up to 690 m GEL have been found for the Paleo-ocean in some geomorphological studies (Carr & Head, 2003; Di Achille & Hynek, 2010). Using these larger estimates for the Paleo-ocean gives a total ice volume ($q_{i,e} + q_i$) that is distinctly larger than the PLD, which indicates that the excess ice could be in the ground.

Second, the diffusion-limited case ($f$ = 0.29–0.33) was considered (Fig. 7c). The interactive evolution of multiple water reservoirs yielded an asymptotic upper limit of the $\delta D_{i,e}$ value (Fig. 6) of 5000‰, which corresponds to the asymptotic upper limit for $f$ = 0.33, $\alpha$ = 1, and $\delta D_i$ = 1000‰. If the escaped volume is sufficiently large, $\delta D_{i,e}$ = 5000‰ is satisfied independently from the escape volume. Thus, no constraints on the upper limit of the escape volume are provided. Although the value of the upper limit depends on the parameters, we suggest the possibility that $\delta D_{i,e}$ = 5000‰ is a natural consequence of multi-reservoir behavior if the atmospheric escape is dominated by diffusion. Ground ice does not necessarily exist in such a case.

In summary, our results indicate two different possibilities depending on the dominant escape regime. In the Jeans escape-limited case (*f* = 0.016–0.11), a comparison of our results with the minimum estimates for the Paleo-ocean suggests that the volume of ground ice present is larger than that of the PLD and/or an unknown permeation mechanism may have occurred. Using larger estimates for the Paleo-ocean suggests that a large amount of ice is present in the ground regardless of the exchangeable volume. However, ground ice does not necessarily exist currently on Mars in the diffusion-limited case (*f* = 0.29–0.33). The high $\delta D_{i,e}$ value (≃5000‰) may correspond to an asymptotic upper limit of the D/H ratio caused by the interactive evolution of the multiple reservoirs.

# 5. Discussion

## *5.1. Evolution of water reservoirs through Martian history*

The surficial water of early Mars is expected to have had a homogeneous D/H ratio because of the high degree of mobility of the liquid water that would have been present. In contrast, we assumed that the hydrogen isotope compositions of the water-ice reservoirs heterogeneously evolved more recently (~3.5 Ga to present) because Mars was cold in that period. This scenario was suggested by the results of recent analyses of Martian meteorites (Usui et al., 2015) and *in situ* measurements of Hesperian mudstone (Mahaffy et al., 2015). Only the topmost ice layer can currently act as a reservoir exchangeable with the atmosphere.

A summary of the evolution of the D/H ratios of the Martian near-surface water reservoirs and the evolution of the total volume through Martian history, calculated by using the multi-reservoir model presented here and a previously published one-reservoir model (Kurokawa et al., 2014), is shown in Fig. 8. The volume was calculated from the volume of the Paleo-ocean (100–690 m GEL) and the δD value (1000–2000‰). The current volume was calculated by using the multi-reservoir model assuming that *f* = 0.016–0.33, and the initial volume was calculated by using the one-reservoir model (Kurokawa et al., 2014).

Two different scenarios are proposed depending on the dominant escape regime. If the hydrogen loss is Jeans escape-limited (*f* = 0.016–0.11), the surficial D/H ratio ($\delta D_{i,e}$ ≃ 5000‰) is satisfied by a relatively small amount of escape. Thus, a large amount of ground ice may currently exist on Mars. If the atmospheric escape is diffusion-limited (*f* = 0.29–0.33), the surficial D/H ratio may be a result of the asymptotic upper limit of the surficial D/H ratio. In this case, a large amount of ground ice does not necessarily exist on Mars today. Therefore, it is crucial to constrain the dominant escape regime and the fractionation factor by detailed modeling of the escape processes.

## *5.2. Possible permeation mechanisms*

Hydrothermal activity, volcanism, and impacts are possible mechanisms of permeation in addition to molecular diffusion. From the geological record, Craddock & Greeley (2009) estimated the minimum volume of water degassed between the Noachian and Amazonian periods to be $\simeq 1 \times 10^{18}$ kg ($\simeq$10 m GEL). Mixing of the outgassed water would contribute to the permeable volume of this amount. Degassing during the pre-Noachian period might have had a more significant impact, although this cannot be estimated from the geological record.

Impacts of asteroids and comets induce hydrothermal activity, which contributes to the mixing of surface water ice. Newsom et al. (1980) estimated that the total volume of impact melt sheets on Mars is greater than 60 m GEL and that the maximum ratio of total steam to melt sheet is 23%. The permeable volume caused by this hydrothermal activity can be estimated as $\simeq$10 m GEL. The volume might be greater if hydrothermal circulation is considered.

The young age of the PLD, $10^{-1}$–$10^{2}$ Ma, estimated from its morphology (Christensen, 2006), suggests that the PLD actively sublimated or melted within the geological timescale. This would contribute to the mixing of the hydrogen isotope composition of the PLD. The discovery of recurring slope lineae in many regions (McEwen et al., 2014) may indicate that partial melting of the ground-ice sheet may induce effective permeation.

## *5.3. Validity of the estimate of escape-induced fractionation factor*

Jeans escape is assumed for the escape regime of hydrogen and deuterium to estimate the fractionation factor induced by atmospheric escape at 3.5 Ga. The validity of the assumption is discussed in this subsection. Because the escape parameter $\lambda$ at the exobase would be sufficiently small for hydrogen ($\lambda$ = 1.5) and deuterium $\lambda$ = 3.0, the escape regime might be the hydrodynamic regime if hydrogen and deuterium are the dominant species at the exobase. However, these species are relatively minor at the exobase in the case considered in this study. We assumed the escape rate to be $\Phi = 5.4 \times 10^{27}$ s$^{-1}$ at 3.5 Ga (Lammer et al., 2003a) in our minimum escape model. Given the Jeans escape velocity of hydrogen $w_J$(H) = $5.8 \times 10^4$ cm s$^{-1}$ (Eq. 13), the number density of hydrogen at the exobase is estimated to be $8.8 \times 10^4$ cm$^{-3}$. In comparison, the total number density at the exobase is $\sim 10^9$ cm$^{-3}$ (e.g., Yung et al., 1988). Thus, hydrogen is a minor component at the exobase even for an ~ 10 times larger escape rate. The thermal escape regime of a $CO_2$-dominated atmosphere after 3.8 Ga is a Jeans escape rather than a hydrodynamic escape (Tian et al., 2009). Therefore, the Jeans escape is valid in the case considered in the present study.

**6. Conclusions**

We calculated the evolution of the reservoir volumes and the D/H ratios by using a multi-reservoir model that considered atmospheric escape and sublimation. The D/H ratio of the topmost exchangeable ice is determined to be comparable to or slightly higher than the atmospheric D/H ratio owing to sublimation-induced isotope fractionation. This result supports the inference that the intermediate-δD reservoir is unexchangeable ice in the ground. We found that the escape regime is a dominant factor in determining a possible scenario. If the hydrogen loss is Jeans escape-limited, a comparison of our results with the geomorphological estimates of Paleo-oceans suggests that the topmost high D/H ice is mixed with a certain volume of subsurface ice over geological timescales and/or that the total volume of ground ice is larger than that of the polar layered ice deposits. However, if the atmospheric escape is diffusion-limited, the observed surficial D/H ratio, which is approximately six times that of the ocean water on Earth, may correspond to the asymptotic upper limit caused by the interactive evolution of multiple water reservoirs. In such a case, a large volume of ground ice does not necessarily exist currently on Mars.

# Acknowledgments

We thank Kosuke Kurosawa for input on asteroid impacts and the members of Mars Science Team of Tokyo Tech for fruitful discussions. This manuscript benefited from constructive comments from anonymous reviewers and the editor, Audrey Bouvier. This work was supported by Grants-in-Aid from the Japanese Ministry of Education, Culture, Sports, Science and Technology (MEXT) (23244027, 23103005 and 26800272). HK was supported by JSPS KAKENHI (Grant No. 15J09448).

Table 1: Martian meteorites and their hydrogen isotope ratios.

| Meteorite | Date | Phase | δD [‰] |
|---|---|---|---|
| Shergotty | 0.17 Ga[a] | apatite | 4600[f] |
| Los Angeles | 0.17 Ga[a] | apatite | 4300[f] |
| GRV 020090 | 0.19 Ga[b] | melt inclusion | 6000[g] |
| LAR 06319 | 0.19 Ga[c] | melt inclusion | 5100[h] |
| LAR 06319 | 0.19 Ga[c] | groundmass glass | 2100[i] |
| EETA79001 | 0.17 Ga[a] | groundmass glass | 1500[i] |
| Y-980459 | 0.47 Ga[d] | groundmass glass | 1600[i] |
| Y-980459 | 0.47 Ga[d] | melt inclusion | 275[h] |
| ALH 84001 | 4.1 Ga[e] | carbonate | 1200[j] |
| ALH 84001 | 4.1 Ga[e] | apatite | 3000[f] |

[a]Nyquist et al. (2001)

[b]Jiang & Hsu (2012)

[c]Shafer et al. (2010)

[d]Shih et al. (2005)

[e]Bouvier et al. (2009); Lapen et al. (2010)

[f]Greenwood et al. (2008)

[g]Hu et al. (2014)

[h]Usui et al. (2012)

[i]Usui et al. (2015)

[j]Boctor et al. (2003)

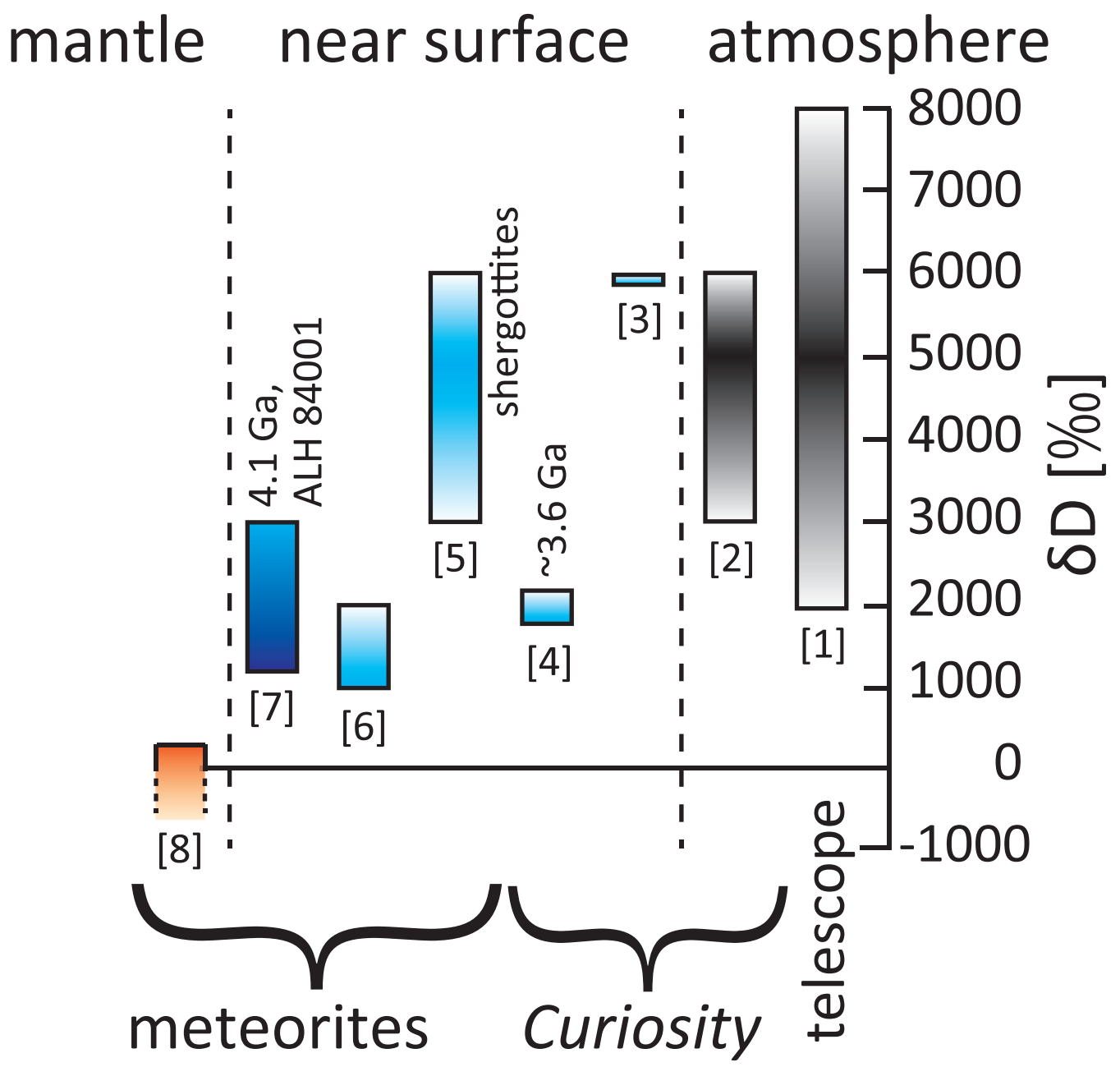

**Figure 1:** Hydrogen isotope compositions of Martian water reservoirs including the atmosphere, near-surface ice, and mantle. The data are from [1] telescope observations of atmospheric deuterium hydrogen isotope ratios (D/H; Owen et al., 1988; Krasnopolsky et al., 1997; Novak et al., 2011; Villanueva et al., 2015; Krasnopolsky, 2015; Aoki et al., 2015), [2] measurements of the atmosphere and [3] soil by the *Curiosity* rover (Webster et al., 2013), [4] measurements of ~3.6 Ga mudstone by the *Curiosity* rover (Mahaffy et al., 2015), [5] analyses of melt inclusions in enriched shergottites originating in the crust (Usui et al., 2012; Hu et al., 2014), [6] analyses of matrix glasses in Martian meteorites suggestive of an independent subsurface hydrogen reservoir as ground ice or hydrated crust (Usui et al., 2015), [7] analyses of carbonate and apatite in ALH 84001 (Boctor et al., 2003; Greenwood et al., 2008), and [8] analyses of melt inclusions in a depleted Martian meteorite originating in the mantle (Usui et al., 2012). Bars denote errors or data ranges.

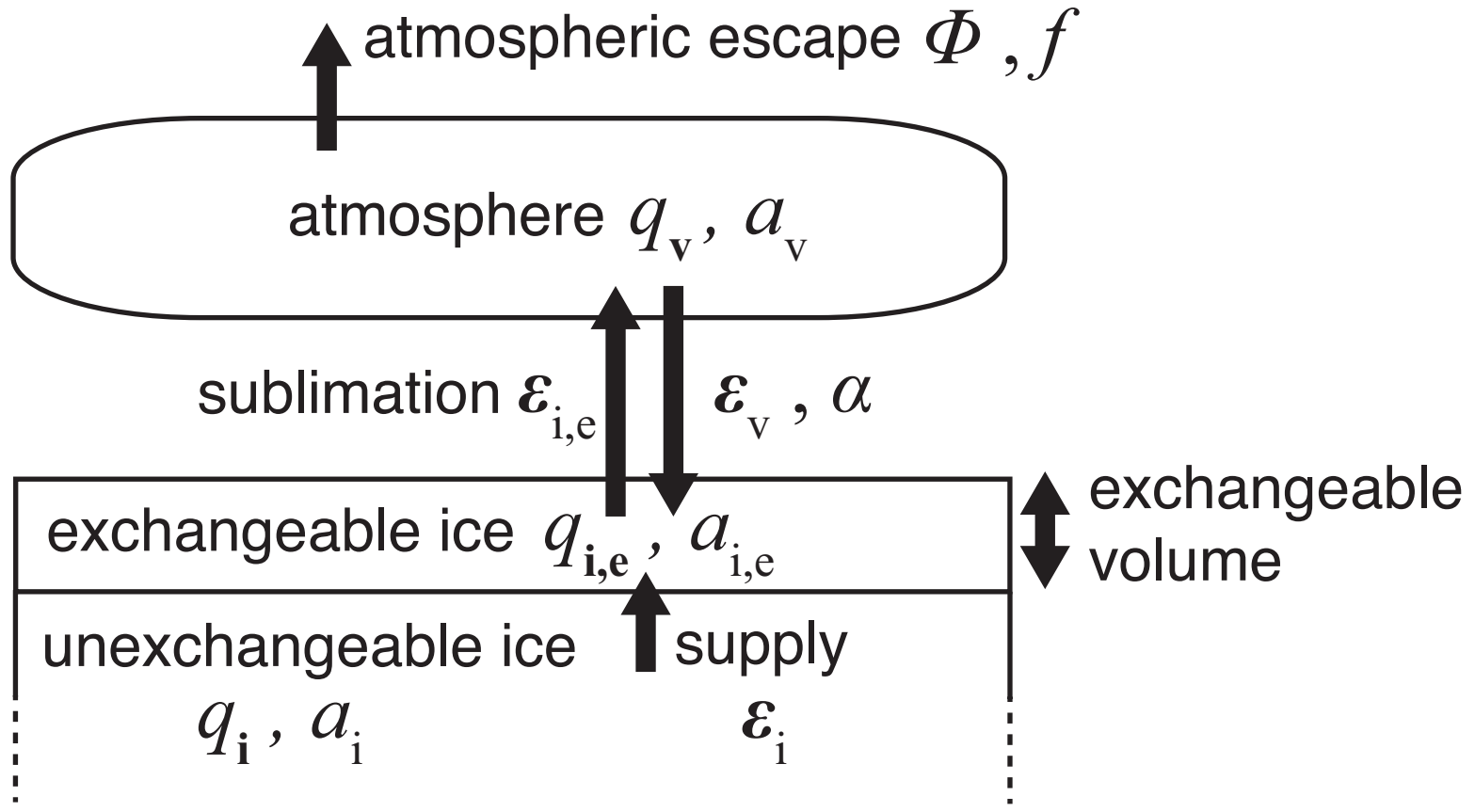

**Figure 2:** Schematic diagram of our model of the evolution of near-surface water reservoirs through the cold stages on Mars.

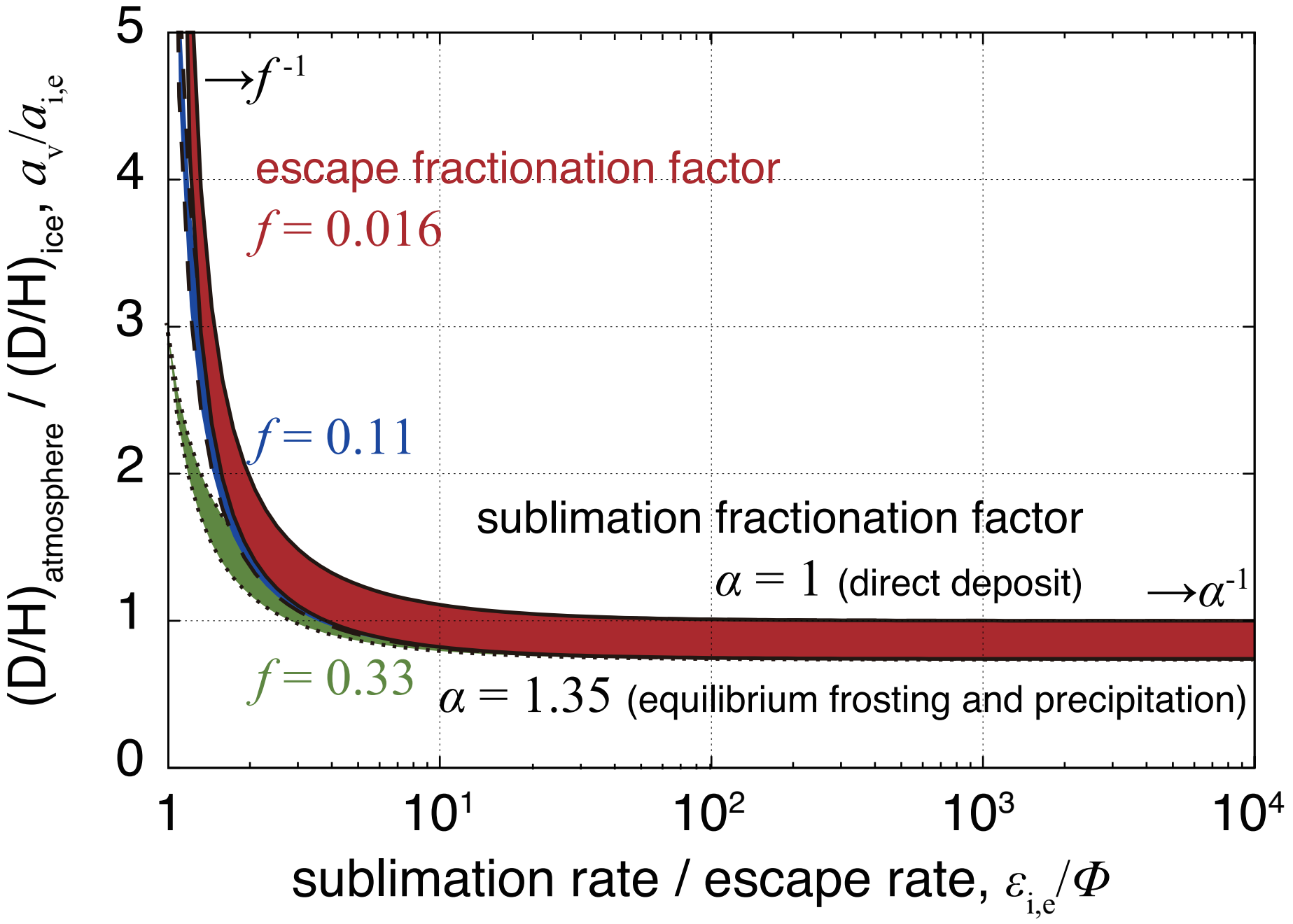

**Figure 3:** D/H ratios of the water in the atmosphere and exchangeable ice ($a_V/a_{i,e}$) as functions of the ratio of the sublimation rate to the escape rate ($\varepsilon_{i,e}/\Phi$) calculated by using our steady state model. The escape fractionation factor is $f = 0.016$ (red, solid lines) for the current conditions on Mars, $f = 0.11$ (blue, dashed lines) for the escape-limited case at 3.5 Ga, and $f = 0.33$ (green, dotted lines) for the diffusion-limited case at 3.5 Ga. The range of each solution (the widths of the red, blue, and green lines) corresponds to the range of the sublimation fractionation factor $\alpha$ = 1–1.35 obtained for direct deposition or frosting and precipitation under equilibrium conditions.

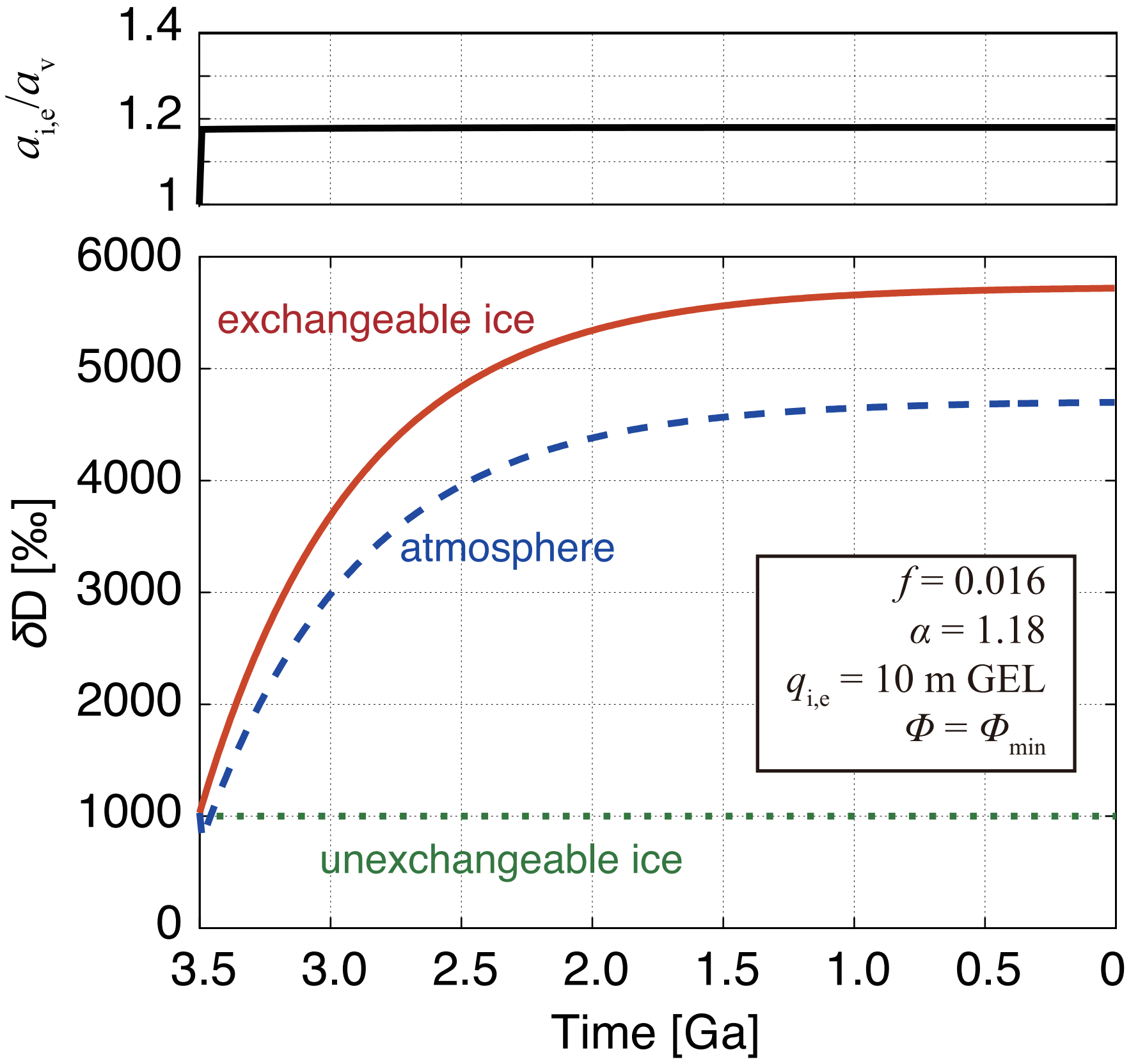

**Figure 4:** Lower panel: Evolution of hydrogen isotope ratios of water in the atmosphere (blue, dashed line), exchangeable ice (red, solid line), and unexchangeable ice (green, dotted line) for $f = 0.016$, $\alpha = 1.18$, $q_{\mathrm{i,e}} = 10$ m GEL, and $\Phi = \Phi_{\mathrm{min}}$. Upper panel: Evolution of the ratio of $a_{\mathrm{i,e}}$ to $a_{\mathrm{v}}$.

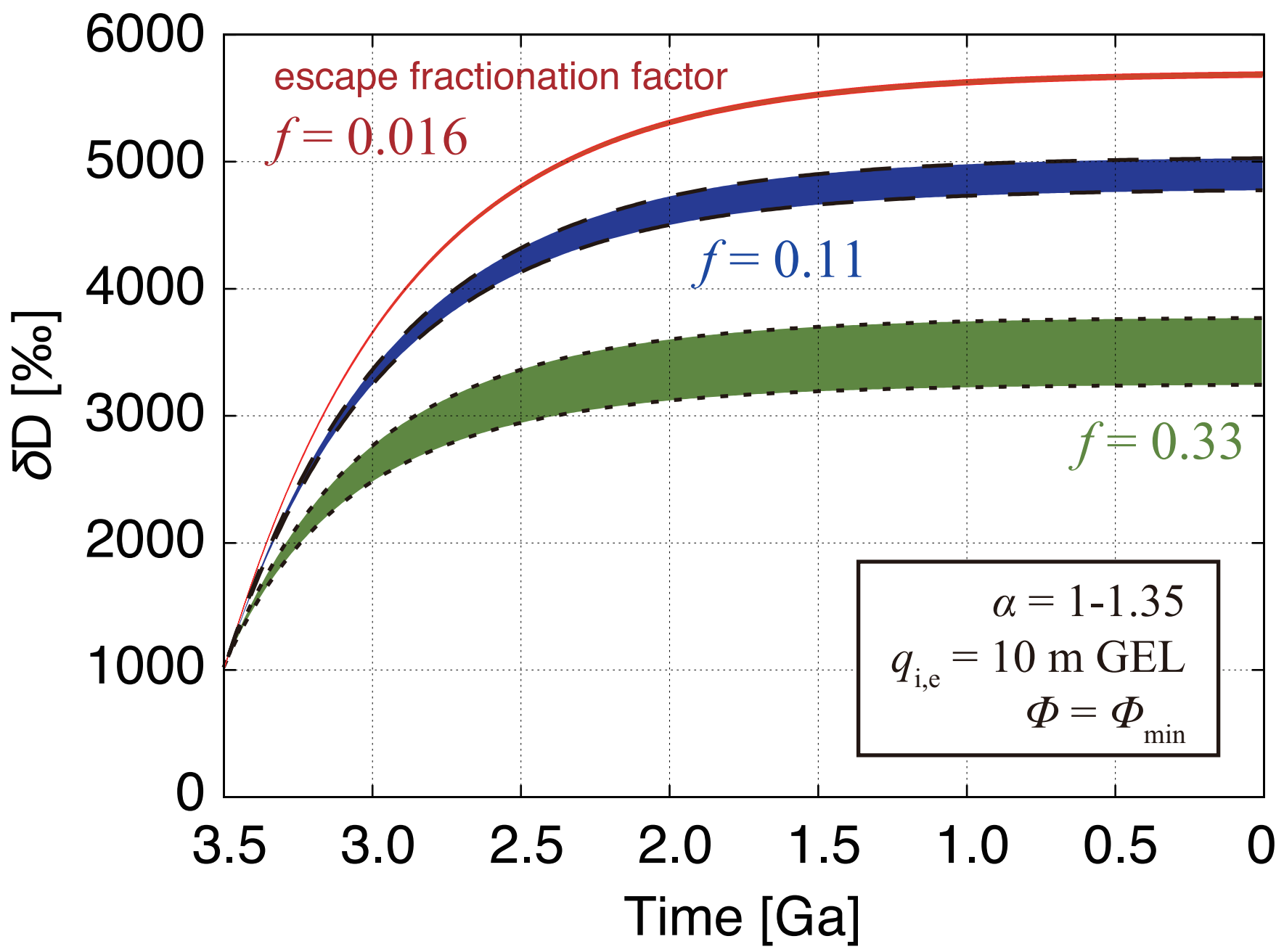

**Figure 5:** Dependence of the evolution of the hydrogen isotope ratio of the exchangeable ice on the fractionation factors $f$ and $\alpha$. Results are shown for $f = 0.016$ (red, solid lines), $f = 0.11$ (blue, dashed lines), and $f = 0.33$ (green, dotted lines). The range of each result is dependent on $\alpha$; a higher D/H is found at a lower $\alpha$.

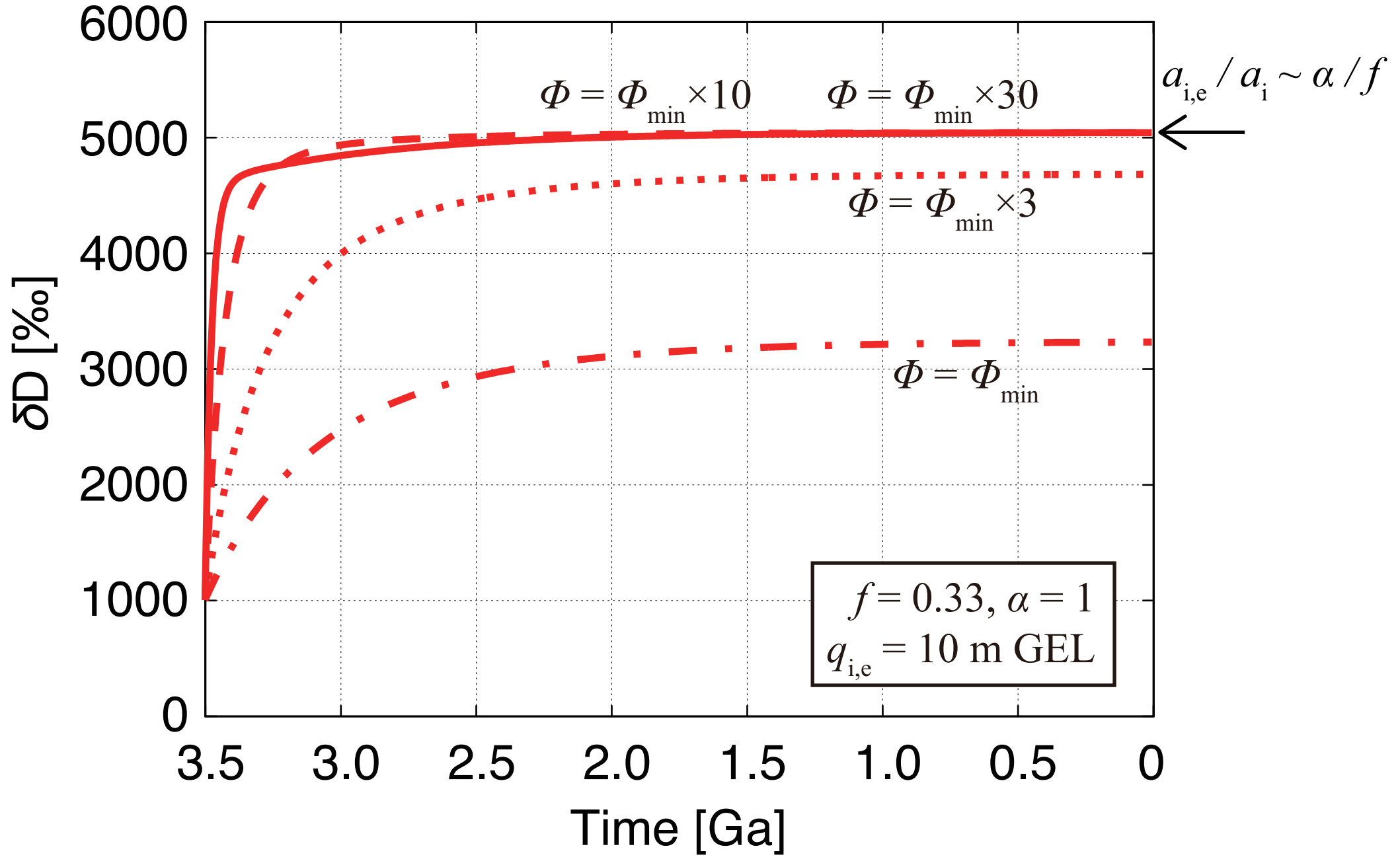

**Figure 6:** Asymptotic behavior of the evolution of the hydrogen isotope ratio of the exchangeable ice for $f = 0.33$, $\alpha = 1$, and $q_{i,e} = 10$ m GEL. The escape rates are $\Phi = \Phi_{min} \times 1$ (dashed and dotted lines), $\Phi_{min} \times 3$ (dotted lines), $\Phi_{min} \times 10$ (dashed lines), and $\Phi_{min} \times 30$ (solid lines).

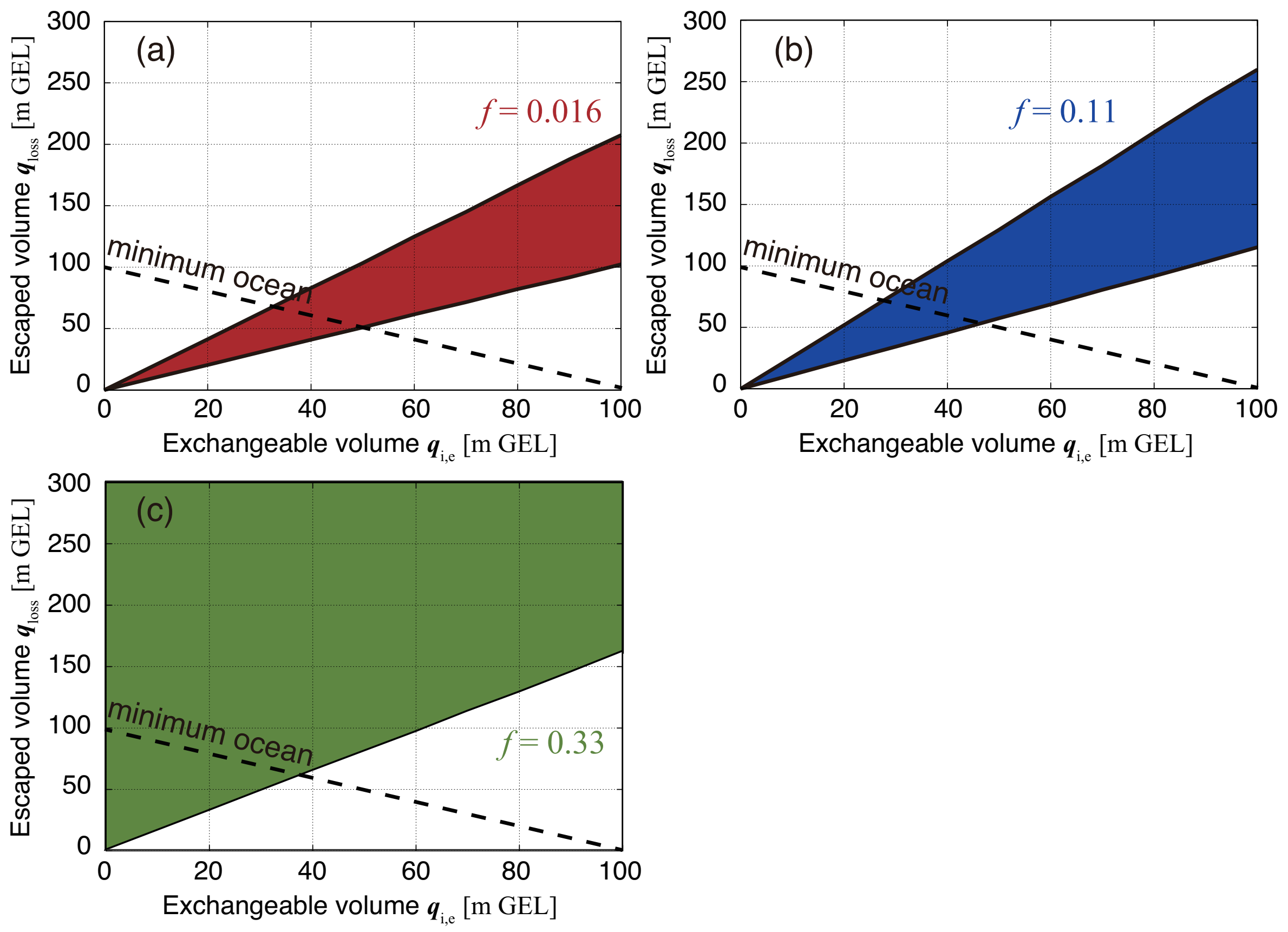

**Figure 7:** Escaped volume required to give δD = 5000‰ for the exchangeable ice as a function of the assumed exchangeable volume. Results are shown for (a) $f = 0.016$, (b) $f = 0.11$, and (c) $f = 0.33$. In each panel, the required water loss is plotted along the vertical axis as a function of the exchangeable volume plotted along the horizontal axis. The range of the escaped volume along the vertical axis is caused by the ranges of the initial δD (1000–2000‰) and $\alpha$ (= 1–1.35). The dashed lines are the minimum of geomorphological estimates for the Paleo-oceans (Head et al., 1999; Carr & Head, 2003; Di Achille & Hynek, 2010).

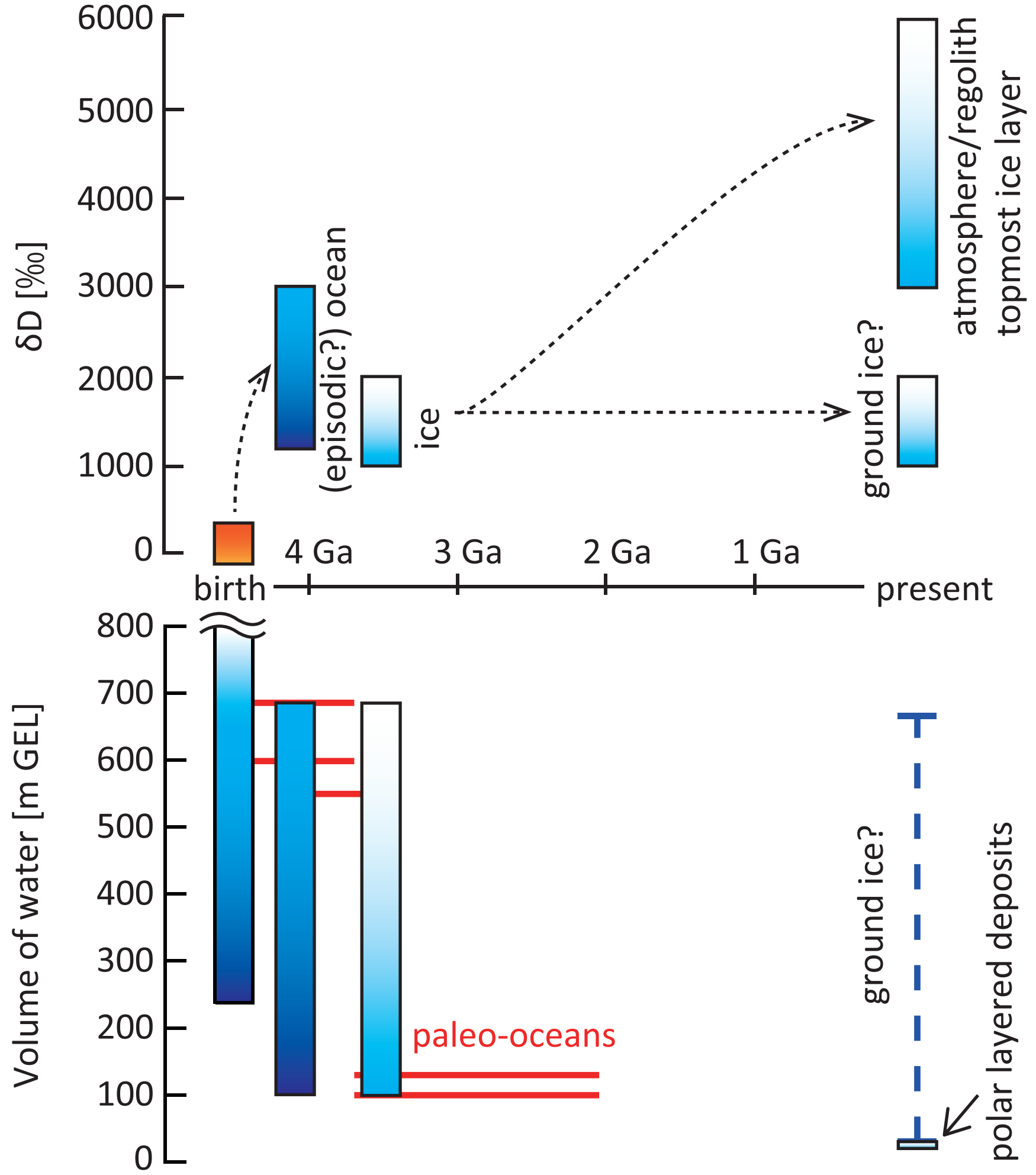

**Figure 8:** Upper panel: Evolution of the hydrogen isotope compositions of Martian water reservoirs. The data sources are the same as those used Fig. 1. Lower panel: Evolution of the volume of the Martian near-surface water. The data used are geomorphological estimates for the Paleo-oceans (Head et al., 1999; Carr & Head, 2003; Di Achille & Hynek, 2010) and the volume of the polar layered deposits (PLD; Zuber et al., 1998; Plaut et al., 2007).